\documentclass[12pt]{article}
\let\section=\subsection     \let\subsection=\subsubsection                
\usepackage{graphicx}
\usepackage{epsfig}
\usepackage{amssymb}
\usepackage{amsmath}
\newcommand{\beq}{\begin{equation}}
\newcommand{\eeq}{\end{equation}}
\newcommand{\bea}{\begin{eqnarray}}
\newcommand{\eea}{\end{eqnarray}}

\newcommand{\ave}[1]{\langle {#1} \rangle}

\newcommand{\pb}{\bar\psi}

\newcommand{\eq}[1]{Eq.~(\ref{#1})}

\newcommand{\ct}{s_{22}}
\newcommand{\cf}{s_{55}}
\newcommand{\cs}{s_{77}}

\def\roughly#1{\mathrel{\raise.3ex\hbox{$#1$\kern-.75em%
\lower1ex\hbox{$\sim$}}}}

\def\={\;=\;}
\def\+{\;+\;}

\begin{document}
\begin{center}
   {\large \bf COLOR-FLAVOR (UN-) LOCKING}\\[5mm]
   M.~OERTEL$^1$ and M.~BUBALLA$^{2,3}$\\[5mm]
   {\small \it  $^1$ IPN Lyon;  43, bv
   du 11 novembre 1918; F-69622 Villeurbanne C\'edex, France \\
   $^2$ IKP, TU Darmstadt, Schlossgartenstr. 9,
   D-64289 Darmstadt, Germany \\
   $^3$GSI Darmstadt;
   Postfach 110552; D-64220 Darmstadt; Germany \\[8mm] }
\end{center}
\begin{abstract}\noindent
The structure of the phase diagram of strongly interacting matter at
moderate densities is calculated within a 3-flavor NJL-type quark
model with realistic quark masses. We focus on the influence of the
selfconsistently determined effective strange quark mass on the
color-flavor unlocking phase transition.
\end{abstract}

\section{Introduction}
\label{introduction}
The structure of the QCD phase diagram as a function of temperature
and density has been studied extensively throughout the last years, on
theoretical as well as on experimantal grounds. For a long time the
discussion has been restricted to two phases: the hadronic phase and
the quark-gluon-plasma. This has
dramatically changed after the discovery that, already for
densities present in the interior of neutron stars, color
superconducting phases might exist with energy gaps of the order of
$\Delta \sim 100$~MeV~\cite{RaWi00}.

At these moderate densities QCD cannot be treated perturbatively and 
therefore one has to rely on model calculations. In general local two-quark
interactions are chosen which are motivated, e.g., from instanton 
interactions. Most interactions favor a condensation in the scalar 
color-$\bar{3}$ channel:
\beq
    s_{AA'} 
    \= \ave{\psi^T \,C \gamma_5 \,\tau_A \,\lambda_{A'} \,\psi} \;.
\label{saa}
\end{equation}
Here both, $\tau_A$ and $\lambda_{A'}$ are the antisymmetric
generators of $SU(3)$, i.e., the antisymmetric Gell-Mann
matrices ($A, A'\; \in\; \{2,5,7\}$), acting in flavor and color 
space, respectively. 

Let us first discuss two limiting cases: Assuming the mass of the
strange quark to be infinite, we can neglect pairing between strange
and lighter quarks and the flavor index in \eq{saa} is
restricted to  $A=2$.
The three condensates $s_{2A'}$ form a vector in 
color space, which always can be rotated into the $A' = 2$-direction. 
Hence the two-flavor superconducting state (2SC) state can be 
characterized by $\ct \neq 0$ and $s_{AA'} = 0$ if $(A,A') \neq (2,2)$.
In this state color $SU(3)$ is spontaneously broken down
to $SU(2)$. For massless $u$ and $d$ quarks the 2SC state
is invariant under chiral $SU(2)_L \times SU(2)_R$ transformations. 

The second limiting case is to assume three degenerate quark
flavors. Then dense matter is expected to form a so-called color-flavor 
locked (CFL) state~\cite{ARW99}, characterized by the situation
$\ct = \cf = \cs \neq 0$ and  $s_{AA'} = 0$ if $A\neq A'$ \,. 
In this state $SU(3)_{color}\times SU(3)_L \times SU(3)_R$
and the $U(1)$-symmetry related to baryon-number conservation are broken
down to a common $SU(3)_{color + V}$ subgroup where color and flavor 
rotations are locked.

Both situations discussed above are idealizations of the real world.
For sufficiently large quark chemical potentials $\mu \gg M_s$, the $s$
quark mass becomes of course almost negligible against $\mu$ and matter 
is expected to be in the CFL phase.  
It is not clear, however, whether this CFL phase is directly connected to
the hadronic phase at low-densities, or whether an intermediate 2SC phase exist, where 
only up and down quarks are paired.  
It is obvious that the answer to this question depends on the strange quark 
mass which is in general $T$- and $\mu$-dependent. Moreover, it could
depend on the presence of quark-antiquark and diquark condensates.
This means, not only the phase structure depends on the 
quark mass, but also the quark mass depends on the phase.  

To study these interdependencies we will use an NJL-type model, where 
the masses are closely related 
to the ($T$- and $\mu$-dependent) $\bar{q}q$ condensates~\cite{hatsuda}
\beq
    \phi_u \= \ave{\bar u u} \= \ave{\bar d d}
    \quad\text{and}\quad
    \phi_s \= \ave{\bar s s}\;.
\end{equation}
This work is largely based on Ref.~\cite{BO01}. 
In addition we will discuss the effects of a flavor mixing interaction, 
which has not been included in that reference.

\section{Model}
\label{model}
We consider the effective Lagrangian
\beq
    {\cal L}_{eff} \= \pb (i \partial\hspace{-2.3mm}/ - \hat{m}) \psi
                      \+ {\cal L}_{q\bar q} \+ {\cal L}_{qq} \,,
\label{Lagrange}
\end{equation}
where $\psi$ denotes a quark field with three flavors and three colors. 
The mass matrix $\hat m$ has the form 
$\hat m = diag(m_u, m_d, m_s)$ in flavor space.
Throughout this article we assume isospin symmetry, $m_u = m_d$.

To study the 
interplay between the color-superconducting
diquark condensates $s_{AA'}$ and the quark-antiquark condensates
$\phi_u$ and $\phi_s$ we consider an NJL-type interaction 
with a quark-quark part
\beq
    {\cal L}_{qq} \=
    H\sum_{A = 2,5,7} \sum_{A' = 2,5,7}
    (\pb \,i\gamma_5 \tau_A \lambda_{A'} \,C\pb^T)
    (\psi^T C \,i\gamma_5 \tau_A \lambda_{A'} \, \psi) 
    \,.
\label{Lqq}
\end{equation}
and a quark-antiquark part
\begin{alignat}{1}
    {\cal L}_{q\bar q} \= G&\, \Big\{\sum_{a=1}^8 \Big[(\pb \tau_a\psi)^2
    \+ (\pb i\gamma_5 \tau_a \psi)^2\Big] + \frac{2}{3}
\Big[(\pb\psi)^2 + (\pb i \gamma_5\psi)^2 \Big]\Big\}\nonumber \\
     -\, K&\,\Big[{\rm det}_f\Big(\pb(1+\gamma_5)\psi\Big) \,+\
                   {\rm det}_f\Big(\pb(1-\gamma_5)\psi\Big)\Big]\;.
\label{Lqbarq}
\end{alignat}
The latter consists of a $U(3)_L \times
U(3)_R$-symmetric 4-point interaction and a 't~Hooft-type 6-point
interaction which breaks the the $U_A(1)$ symmetry.

Starting from this Lagrangian we calculate the mean-field
thermodynamic potential $\Omega$ at temperature $T$ and quark chemical 
potential $\mu$  and determine the selfconsistent solutions for the 
expectation values $\phi_u=\phi_d$, $\phi_s$, $\ct$, and $\cf=\cs$
by minimizing $\Omega$ with respect to these expectation values.
In this context it is convenient to introduce the effective 
quark masses \mbox{$M_i = m_i - 4G\phi_i + 2 K \phi_j\,\phi_k$,}
($(i,j,k) =$ cyclic permutations of $(u,d,s)$) and the diquark gaps
$\Delta_i = -2H s_{ii}$.

\section{Numerical results}
\label{results}   

To determine the values of the various condensates we first have to
specify the parameters of the interaction.  As
a starting point we take the parameter
values obtained in Ref.~\cite{Rehberg} by fitting vacuum masses and
decay constants of pseudoscalar mesons. It is, however, certainly
worthwhile to study in detail the dependence of the results on the
strength of the 't Hooft-type interaction which mixes different
flavors and thus affects $M_s$ even in regimes, where
no strange quark states are populated. To that end we vary 
the coupling constant $K$ and readjust $G$ and $m_s$ such that the vacuum 
masses stay unchanged. Since the quark-quark part of the
interaction does not influence any vacuum properties the coupling
constant $H$ has to be determined in a different way. 
For $K=0$ we take $H = 3/4 G$ which corresponds to the ratio of $G$ and $H$
obtained from a four-fermion interaction with the quantum numbers of a 
single-gluon exchange~\cite{ARW99,ABR99}. The resulting value for $H$ is
then kept constant for all values of $K$. 
As in Ref.~\cite{Rehberg}, all quark loops are regularized by a sharp 
3-momentum cut-off $\Lambda$. 

\begin{figure}
\begin{center}
\epsfig{file=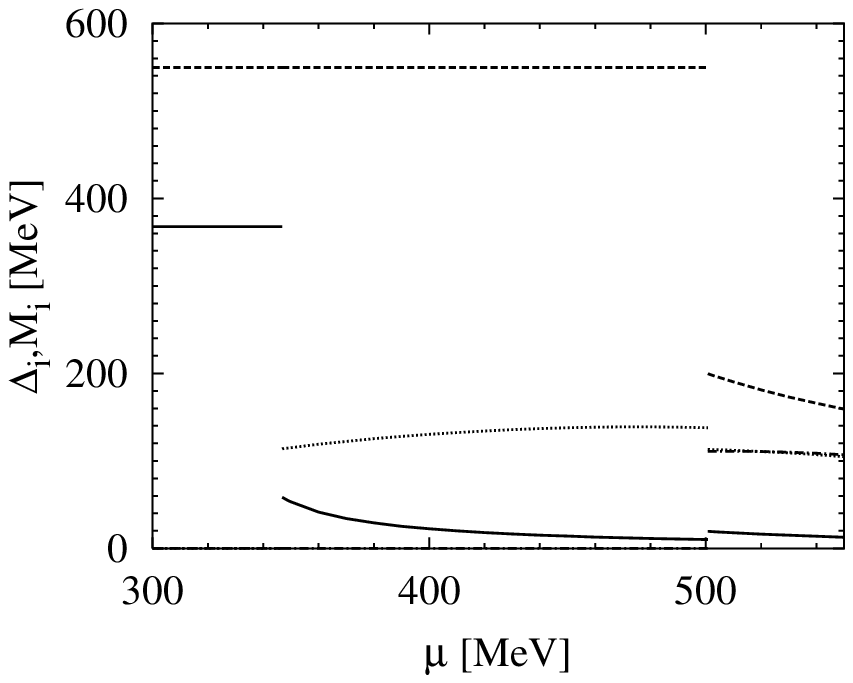,width=5.5cm}
\hfill
\epsfig{file=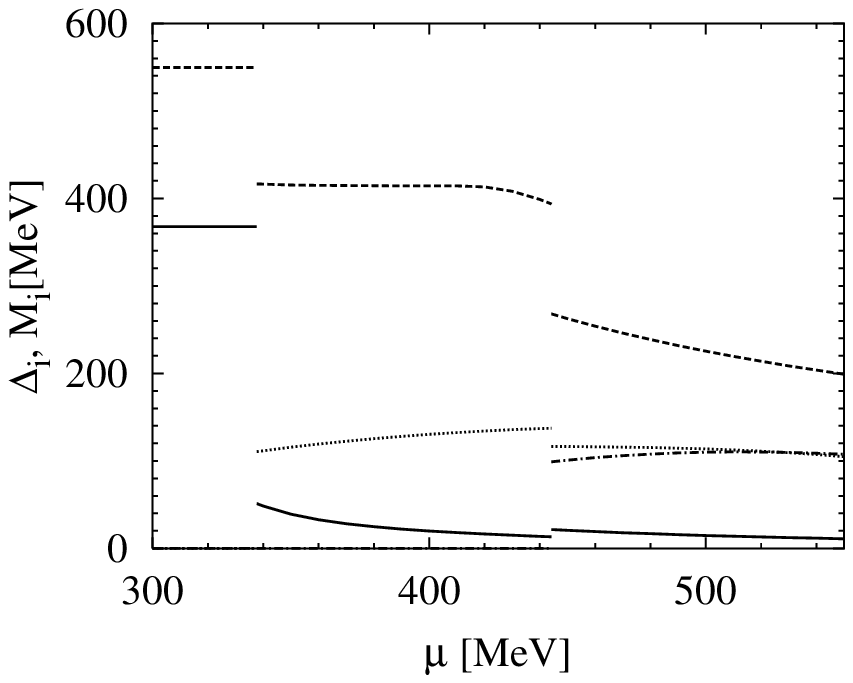,width=5.5cm}
\end{center}
\vspace{-0.5cm}
\caption{\small Gap parameters at $T=0$ as functions of $\mu$:
                Constituent masses $M_u=M_d$ (solid) and $M_s$ (dashed),
                and diquark gaps $\Delta_2$ (dotted) and $\Delta_5$
                (dashed-dotted). Left: $K = 0$.
                Right: $K \Lambda^5 = 20$.}
\label{figt=0}
\end{figure}

We begin with the discussion of the results at $T=0$.  
The behavior of the constituent quark masses and the
diquark gaps as functions of $\mu$ is displayed 
in Fig.~\ref{figt=0}. The left panel corresponds to $K=0$, the right panel 
to $K\Lambda^5 = 20$. One can clearly distinguish three phases. 
At low $\mu$, the diquark gaps 
vanish and the constituent quark masses stay at their vacuum values.
Hence, in a very schematic sense, this
phase can be identified with the ``hadronic phase''.

At a critical chemical potential $\mu = \mu_1$ a first-order phase 
transition to the 2SC phase takes place: The diquark
gap $\Delta_2$ has now a non-vanishing value, whereas $\Delta_5$
remains zero. At the same time the mass of the light quarks drops from
the vacuum value to about 50~MeV and the baryon number density jumps
from zero to about 2.5 nuclear matter density. This behavior,
including the value of $\mu_1$, is rather independent of $K$. Obviously
this is not the case for $M_s$: For $K = 0$ the strange quarks
are completely unaffected, whereas a nonzero value of $K$ leads to a
considerably lower strange quark mass in the 2SC phase. This favors
the population of strange quark states already in the 2SC phase and
indeed we observe a nonzero density of strange quarks above $\mu
\approx 415$~MeV. This is the reason for the accelerated decrease of
$M_s$ which can be observed in Fig.~\ref{figt=0}.

At $\mu = \mu_2$ the system undergoes a second first-order phase
transition, this time from the 2SC phase into the CFL phase, which is
characterized by a non-vanishing diquark gap $\Delta_5$ (together with
a non-vanishing $\Delta_2$). The flavor-mixing interaction accelerates
the transition to the CFL phase: The value of $\mu_2$ for $K \Lambda^5
= 20$ is about 50~MeV lower than for $K  = 0$. 
On the other hand, the values of $\Delta_5$ and $\Delta_2$ for a given
value of $\mu$ depend only weakly on $K$, as long as the CFL solution
is stable. Since in both cases there still is a metastable CFL solution
below $\mu_2$, the main effect of the flavor mixing interaction is
to render the CFL solution stable for lower values of $\mu$.

\begin{figure}
\begin{center}
\epsfig{file=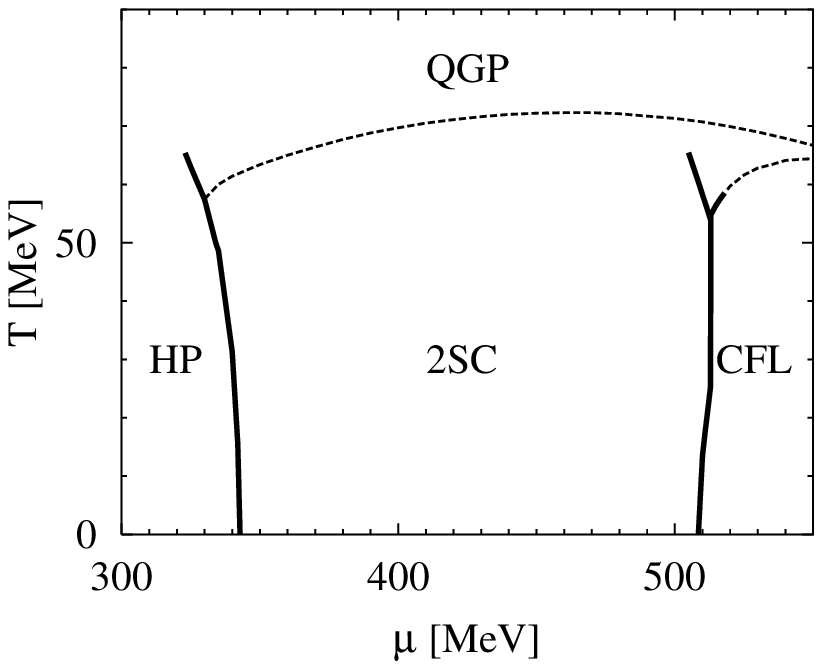,width=5.5cm}
\hfill
\epsfig{file=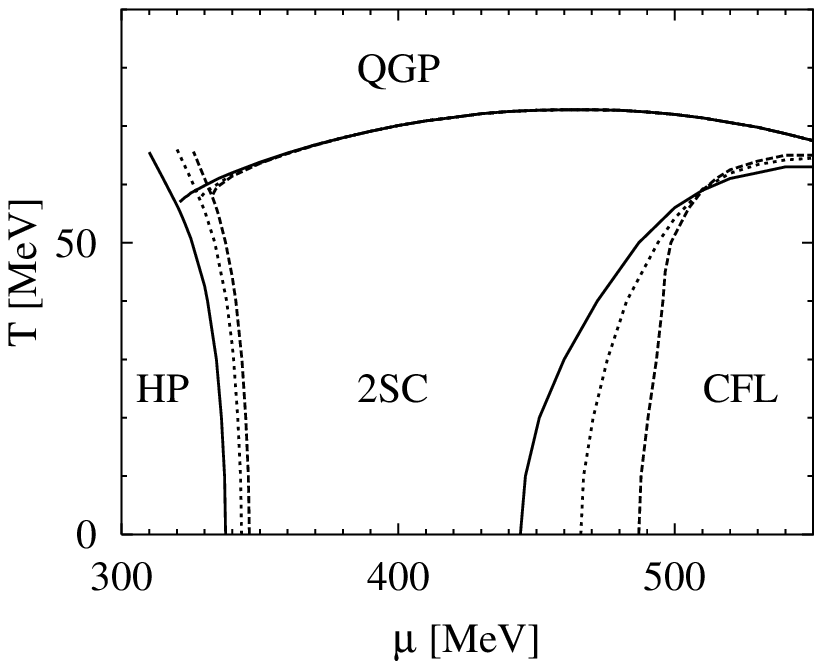,width=5.5cm}
\end{center}
\vspace{-0.5cm}
\caption{\small Left: Phase diagram for $K  = 0$ (from \cite{BO01}). 
         The lines indicate phase boundaries with first-order (solid)
         and second-order (dashed) phase transitions.
         Right: Phase diagram for $K \Lambda^5 = 5$ (dashed), 12.36
         (dotted), and 20 (solid).}
\label{figphase}
\end{figure}

We now extend our analysis to non-vanishing temperatures. The
resulting phase diagrams in the $\mu$-$T$ plane are shown in
Fig.~\ref{figphase}. The left panel corresponds to the results of 
Ref.~\cite{BO01}, obtained with a slightly different parameter set with 
$K=0$. We can distinguish four different regimes: At
low temperatures we find (with increasing $\mu$) the
``hadronic phase'', the 2SC phase, and the CFL phase.  Similar to zero
temperature these three phases are well separated by first-order phase
transitions.  The high temperature regime is governed by the QGP
phase, which is characterized by vanishing diquark condensates and
small values of $\phi_u$ and (for large enough $\mu$) $\phi_s$.  There
we find smooth crossovers with respect to $\phi_u$ and $\phi_s$
instead of the first-order phase transitions, which therefore end in
second order endpoints. 
The transition from
the 2SC phase to the QGP phase is of second order and the critical 
temperature is in almost perfect agreement with the well-known BCS
relation $T_c = 0.57 \Delta_2 (T = 0)$.

Since the diquark gap $\Delta_5$  vanishes at a somewhat lower temperature
than $\Delta_2$, there is an intermediate 2SC phase ``above'' the CFL phase.
It has been argued~\cite{ABR99} that this
color-flavor-unlocking transition has to be first order because
pairing between light and strange quarks can only occur if the gap is
of the same order as the mismatch between the Fermi surfaces. Hence
the value of the gap must be discontineous at the phase transition.
Moreover, the phase transition corresponds to a finite temperature chiral 
restoration phase transition in a three-flavor theory, and therefore the
universality arguments of Ref.~\cite{PiWi84} should apply~\cite{RaWi00}.
At low temperatures our results are in agreement with these
predictions. However, above a critical point we find a second order
unlocking transition. In fact, the above arguments are not as
stringent as they seem to be on a first sight: First of all the Fermi
surfaces are smeared out due to thermal effects and secondly the
universality argument is not rigourously applicable because the 2SC
phase is not a three-flavor chirally restored phase, but only
$SU(2)\times SU(2)$ symmetric. Note that even $SU(3)_V$ is broken
explicitly and spontaneously.
  
On the r.h.s. of Fig.~\ref{figphase} we show first results on the variation 
of the phase diagram with $K$. We find that the 't Hooft interaction mainly
influences the unlocking transition at low $T$. The other
boundaries remain almost unchanged. 

\section{Conclusions}
\label{conclusions}
We studied the ``cold dense'' region of strongly
interacting matter within the framework of an NJL-type quark
model. Our main intention was to examine the effect of
selfconsistently determined realistic
quark masses, in particular of the strange quark mass.
Our results indicate that the 2SC state is the most
favored one up to rather high densities.
The color-flavor unlocking transition is mostly triggered
by a discontinous change of the strange quark mass. Including a flavor
mixing interaction, which lowers the strange quark mass within the 2SC
phase, the CFL phase becomes favored for smaller $\mu$.
With increasing temperature the values of the diquark condensates
decrease and vanish at some critical value $T_c$ which is lower for
$\Delta_5$ than for $\Delta_2$. This color-flavor unlocking transition
becomes second order above a critical point.

For simplicity, we restricted our studies to a common baryon chemical 
potential for all flavors. However, for many applications, 
like quark cores of neutron stars, it would be interesting
to consider $\beta$-equilibrated quark matter, where the chemical
potential for $u$ quarks is in general different from that of $d$ and
$s$ quarks. 
Moreover, since the effective interaction at the densities of interest
is rather unknown, a more systematic examination of the sensitivity of 
the results on the choice of the interaction should be performed to sort 
out some common features.

\end{document}